\documentclass{article} % For LaTeX2e
\usepackage[preprint]{colm2026_conference}

\usepackage[T1]{fontenc}
\usepackage{upquote}

\usepackage{microtype}
\usepackage{hyperref}
\usepackage{url}
\usepackage{booktabs}
\usepackage[table]{xcolor}
\usepackage{multirow}
\usepackage{graphicx}
\usepackage{amsmath}
\usepackage{subcaption}

\usepackage{lineno}

\usepackage{listings}
\definecolor{darkblue}{rgb}{0, 0, 0.5}
\hypersetup{colorlinks=true, citecolor=darkblue, linkcolor=darkblue, urlcolor=darkblue}

\title{Can LLMs Emulate Human Belief Dynamics?}

\author{Adiba Mahbub Proma\\
Department of Computer Science\\
University of Rochester
\And 
Neeley Pate \\
Department of Computer Science\\
University of Rochester
\AND
James N. Druckman \\
Department of Political Science\\
University of Rochester
\And 
Gourab Ghoshal\\
Department of Physics and Astronomy\\
University of Rochester
\AND
Hangfeng He\\
Department of Computer Science\\
University of Rochester
\And 
Ehsan Hoque\\
Department of Computer Science\\
University of Rochester
}

\begin{document}

\ifcolmsubmission
\linenumbers
\fi

\maketitle

\begin{abstract}
Can LLMs simulate how humans form and change beliefs in social networks? We put this to the test by replicating an established study on belief dynamics, evaluating 12 LLMs across multiple model families and parameter sizes. The answer is a clear no, and in systematic ways. LLMs fail to capture initial human belief distributions and tend to be overall more conformist than humans, shifting their responses to align with those around them. They also take a nuanced approach to emulating human homophilic tendencies within networks. Our findings carry a double payoff: they highlight fundamental properties of LLM behavior, and they raise a sharp warning against deploying LLMs as human proxies in social simulations.

%Large Language Models (LLMs) are increasingly used to simulate human attitudes, opinions, and behaviors in large-scale social science surveys and experiments. In this research, we explore whether LLMs can simulate human belief dynamics within network settings by replicating an existing study on belief dynamics (341 participants, 1023 total instances). Evaluating LLM capabilities across 12 different models of various families and parameters, we show that LLM behavior systematically diverges from humans when it comes to belief changes and network restructuring. LLMs tend to be more easily influenced by others' opinions compared to humans, but take a more nuanced approach to emulating homophily. Our findings, therefore, highlight potential pitfalls of using LLMs as human proxies in simulations.

\end{abstract}

\section{Introduction}
\vspace{-0.5em}
Large Language Models (LLMs) are increasingly used to simulate human attitudes, opinions and behaviors \citep{Argyle_2023, park2024generative}. Such endeavors include but are not limited to the replication of survey outcomes \citep{ Argyle_2023, Bisbee_Clinton_Dorff_Kenkel_Larson_2024}, social interactions \citep{Park_OBrien_Cai_Morris_Liang_Bernstein_2023, park2024generative}, and reasoning trends \citep{pate2026replicating}. This can be highly advantageous for research where there are ethical concerns such as humans being exposed to harmful content, hate speech or misinformation, or even when there are resource constraints (as getting data from humans is expensive). Yet, there are concerns about how accurately LLMs can represent human tendencies, particularly in network settings. 

One emerging area of simulation is changes in beliefs and opinion dynamics \citep{Chuang_Goyal_Harlalka_Suresh_Hawkins_Yang_Shah_Hu_Rogers_2024,Chuang_Nirunwiroj_Studdiford_Goyal_Frigo_Yang_Shah_Hu_Rogers_2024}. Prior works have explored how LLMs can be used to simulate network-level interactions \citep{Park_OBrien_Cai_Morris_Liang_Bernstein_2023, Chuang_Goyal_Harlalka_Suresh_Hawkins_Yang_Shah_Hu_Rogers_2024}, and found that LLMs can reproduce human behaviors such as organizing a party \citep{Park_OBrien_Cai_Morris_Liang_Bernstein_2023} or reaching scientific consensus \citep{Chuang_Goyal_Harlalka_Suresh_Hawkins_Yang_Shah_Hu_Rogers_2024}. However, there is still limited work on how these LLMs perform when provided with competing opinions and asked to select preferences for opinions. This is highly crucial to explore since they can affect downstream tasks such as simulation experiments related to network tendencies like polarization and persuasion, or for designing interventions. In this research, we extend existing works in simulation of belief dynamics by exploring how well LLMs can simulate belief and network changes. 

A promising approach in LLM simulations is creating ``digital twins'' of real people \citep{toubia2025database, li2025far}. In one research study, LLMs were used to create digital twins of 1000 individuals using information from qualitative interviews \citep{park2024generative}. Simulated agents were able to replicate responses almost as accurately as participants could replicate their own answers two weeks later \citep{park2024generative}, suggesting that ``digital twins'' in simulations might be better able to replicate human tendencies. This also approach allows one-to-one mapping of LLM to human tendencies, which better highlights LLM pitfalls in simulations. However, the personas created in these papers for one-to-one mapping is quite comprehensive, and that level of information may not always be available to researchers. Therefore, it is crucial to explore if the ``digital twins'' method works with simpler personas. 

%Taking that into account, we create digital twins of 1023 participants using a subset of data from a study on human belief dynamics for political topics \citep{proma2025personalized}, 

In this paper, we explore whether LLMs can emulate belief dynamics and network behaviors by simulating an existing study on human belief dynamics for political topics \citep{proma2025personalized}. In the actual study, participants go through three stages - they respond to specific prompts about a topic, are shown the responses of others and can change their own response, and can select who they want in their networks. We use a subset of the data from the study (341 participants, 1023 total samples), and create ``digital twins'' using the demographic and personality-related information of those participants. The ``digital twins'' LLMs are then prompted to go through each of the stages. We aim to answer the following research questions: 
\begin{itemize}
 \vspace{-0.5em}
    \item \textbf{RQ1:} How well can LLMs emulate human beliefs and changes in beliefs? 
     \vspace{-0.5em}
    \item \textbf{RQ2:} How do LLMs differ from humans in choosing which opinions to retain in their networks?
    %Do LLMs, acting as human proxies, selectively amplify certain opinions within their networks?
 \vspace{-0.5em}
\end{itemize}

% Taking that into account, we consider a subset of the data from an existing study on beliefs \citep{proma2025personalized}, and create digital twins of 1023 participants. 

% In this paper, we explore whether LLMs can emulate belief dynamics and network behaviors through simulations using digital twins. We consider a subset of the data from a study on beliefs by Proma et al.(\citeyear{proma2025personalized}), which looked at human belief dynamics within networks for political topics. 

To answer these research questions, we evaluate the replication capabilities of 12 different models from four families and various parameter sizes. We find that LLM behavior systematically diverges from humans when it comes to belief dynamics. LLMs tend to be more easily influenced by others' opinions (i.e., they change their responses to match that of others around them). In terms of network restructuring, LLMs take a nuanced approach, as they can emulate selections made by humans but cannot emulate human homophilic tendencies. Our findings, therefore, highlight potential pitfalls of using LLMs, particularly those with constrained personas, as human proxies in simulations.
\section{Background and Related Works}
\vspace{-0.5em}
\subsection{Theories on human belief dynamics} %Belief change 
\vspace{-0.5em}
Understanding how or why human beliefs change is a well-explored field within psychology and computational social science, with emphasis on several domains such as politics \citep{costello2024durably}, climate change \citep{proma2025exploring}, health \citep{Dalege_VanDerDoes_2022}, and science \citep{Nyhan_Porter_Wood_2022}. Prior work on belief change has emphasized that humans are often rigid, either not updating beliefs based on new information or updating in small increments \citep{proma2025exploring, Introne_2023}. When encountering situations where beliefs could be updated, both the initial state of the individual's beliefs and the position of the new information they're exposed to can play a role. For example, the strength of the arguments presented as well as the perceived credibility of the source delivering the information have been thought to affect how beliefs change \citep{Petty_1986}. When individuals do choose to update their beliefs, scholars have suggested that individuals consider their initial position and the positions of those around them, \citep{degroot1974reaching, weisbuch2002meet}, potentially depending on their trust in others \citep{degroot1974reaching}. Research also suggests that exposure to contradictory viewpoints can ``backfire'', where an individual's perspective becomes reinforced \citep{Bail_Argyle_Brown_Bumpus_Chen_Hunzaker_Lee_Mann_Merhout_Volfovsky_2018, Nyhan_Reifler_2010}. Similarly, seeing counter-attitudinal messages may lead to discounting of the information \citep{Eil_Rao_2011}, leaving initial attitudes unchanged \citep{Druckman_McGrath_2019}. Ultimately, various factors and psychological processes impact human belief dynamics. Considering this complexity, it is therefore crucial to understand whether LLMs can replicate such belief patterns in simulations.

\subsection{LLM-based simulations} %use of LLM based simulations for understanding humans 
\vspace{-0.5em}
Researchers have experimented with using LLMs to replicate certain human behaviors. One popular area of study is survey replication, where LLMs are often given different personalities to help recreate distributions similar to a ``representative sample'' \citep{Zhang_Xu_Alvero_2025, Argyle_2023, Bisbee_Clinton_Dorff_Kenkel_Larson_2024}. Another area of exploration for LLM simulation is motivations and their effect on information processing, such as information evaluation \citep{Dash_Reymond_Spiro_Caliskan_2025}, information summarization \citep{Cho_Hoyle_Hermstrwer_2025}, and opinion formation \citep{pate2026replicating}. These simulations show that LLMs may reach different outcomes based on their provided personas \citep{Dash_Reymond_Spiro_Caliskan_2025, Cho_Hoyle_Hermstrwer_2025}. However, these diverse outcomes may not be reflective of a human population, failing to match the full variation of a sample population \citep{Bisbee_Clinton_Dorff_Kenkel_Larson_2024} and population subgroups \citep{ Zhang_Xu_Alvero_2025}.

% In such simulations, the results are often mixed as to whether LLMs can accurately reflect responses from the demographics they're trying to recreate, such as flattened variation \citep{Bisbee_Clinton_Dorff_Kenkel_Larson_2024, Zhang_Xu_Alvero_2025}. 

Beyond aggregating individual responses into representative samples, LLMs have been used to simulate network-level interactions. For example, providing LLMs with memory and ability to reflect within complex network systems that mimic human everyday lives can help simulate natural human behaviors \citep{Park_OBrien_Cai_Morris_Liang_Bernstein_2023} or replicate individual actions in behavior-based dilemmas \citep{park2024generative}. Similar research has measured how LLM opinions change in one-on-one interactions within a network \citep{Chuang_Goyal_Harlalka_Suresh_Hawkins_Yang_Shah_Hu_Rogers_2024} and how information, such as rumors, spreads in various network configurations \citep{Hu_Liakopoulos_Wei_Marculescu_Yadwadkar_2025}. We extend existing work in LLM belief simulation by adding natural complexities of human belief formation: competing information streams and agency over information stream preference (e.g., ``following'' behavior).

% We extend existing work in opinion formation for LLM simulation within networks by introducing competing information streams that's provided to the LLM, and providing LLMs the opportunity to select what information they prefer. 

%We extend work in opinion formation in network interactions by introducing competing information streams, the opportunity to update networks, and basing LLM information on real human data.

\subsection{LLMs in political domain} %Agent Based Modeling in Political domain/conversational/belief change 
\vspace{-0.5em}

Along with utilizing LLMs for simulating survey responses or networked interactions, LLMs have been incorporated into political experiment design. Prior works show LLMs are capable of incorporating personalization for effective belief formation interventions \citep{Proma_Pate_Druckman_Ghoshal_Hoque_2025, costello2024durably, matz2024potential}. Conversations with LLMs have the potential to even change behavior, for example, by swaying opinions of political candidates in registered voters \citep{Potter_Lai_Kim_Evans_Song_2024}. These studies suggest that LLMs are persuasive argument creators, particularly within the political domain \citep{Hackenburg_Margetts_2024, costello2024durably, proma2025personalized}. We extend work in political experiments by seeing how LLMs are influenced by their environment within a political setting.

% Along with utilizing LLMs for replicating human tendencies such as survey responses or networked interactions, LLMs have also been incorporated into political experiment design. There are several potential applications for LLMs, such as studies of misinformation or political beliefs. Recent work in LLM-based interventions for misinformation include fact-checking \citep{pan2023fact}, misinformation detection \citep{lucasetal2023fighting}, and encouraging critical thinking \citep{Tang_Singha_2024}. More advanced architectures have been constructed around LLMs to incorporate elements of personalization for belief formation interventions \citep{Proma_Pate_Druckman_Ghoshal_Hoque_2025, costello2024durably, matz2024potential}. 

% Other work has explored the potential effects of LLM political biases, which tend to be left-leaning \citep{faulborn-etal-2025-little}. Such preferences can affect news source selection \citep{Proma_Pate_Druckman_Ghoshal_He_Hoque_2025, Dai_Cao_Wang_Pang_Xu_Ng_Chua_2025},  ratings of accuracy based on source leaning \citep{Yang_Menczer_2025}. Conversations with LLMs can sway opinions on political candidates \citep{Potter_Lai_Kim_Evans_Song_2024}. These studies suggest that LLMs are persuasive argument creators, particularly within the political domain \citep{Hackenburg_Margetts_2024, costello2024durably, proma2025personalized}. We extend work in political experiments by seeing how LLMs are influenced by their environment within a political setting.
\section{Methods}
\vspace{-0.5em}
\begin{figure}[t]
\centering
\includegraphics[scale=0.3]{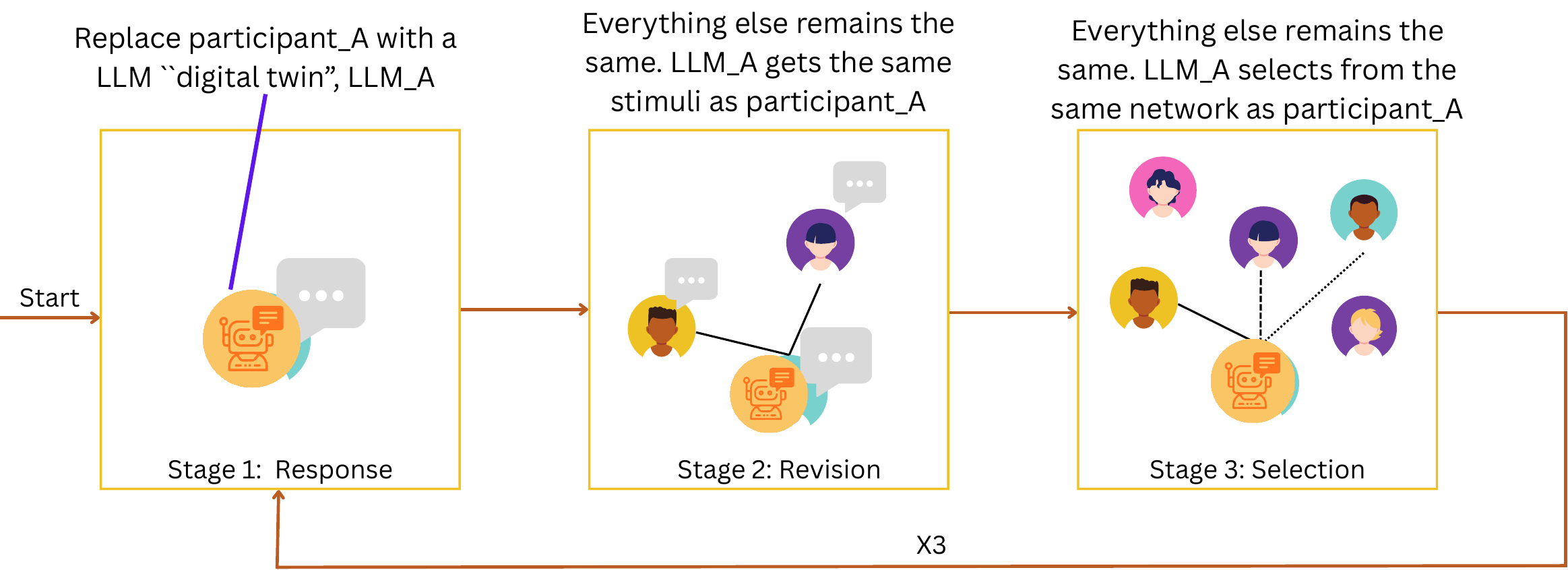}
\caption{Experimental setup for the LLM simulations. In the actual experiments, each participant goes to the three stages meant to understand their initial beliefs, change in beliefs, and network restructuring based on their selection choices. In our experiment, LLMs go through the same set of stages, replacing the participant.}
\label{fig:framework}
\end{figure}

\subsection{Study details} 
\vspace{-0.5em}
In the original study \citep{proma2025personalized}, participants are first presented with a statement (either true or false) on an issue, and they rate how much they believe the statement to be true on a 5-point Likert Scale (stage 1), where 0 is coded as strongly disagree and 4 is coded as strongly agree. They are then exposed to others' belief of the same statement and have the option to update their own beliefs (stage 2). Finally, they are asked to follow/unfollow individuals for subsequent rounds (stage 3). We replicate this exact experimental setup from each player's point of view but with LLMs, including what they were exposed to during the experiment. Following the same structure, we prompt the LLM with the participant's demography and the Big-5 personality traits \citep{rammstedt2007measuring} (i.e., they are given a persona). We then prompt the model to go through each of these stages (see Appendix, section \ref{prompts} for exact prompts). As shown in Figure \ref{fig:framework}, the ``digital twin'' LLM is provided with the same stimuli as a specific participant received in the experiment. While the original study consists of four topics of varying saliency, we focus on two of the four topics: ``immigration'', and ``oil and fuel''. This results in 341 ``digital twins'', and as each participant completed 3 rounds, we had a total of 1023 samples. 

%Immigration reflects a high salience and non-economic issue, overpopulation reflects a low salience and non-economic issue, and fuel and  oil reflect a low salience and economic issue\footnote{This issue was coded as low salience in 2024 when the studies were initially conducted.}.

\subsection{Experimental setup} 
\vspace{-0.5em}
We evaluated LLMs from different model families and of various parameter sizes. We evaluate both non-thinking models (non-reasoning models) and thinking models (reasoning models). We select the following non-thinking models: gemma3\_4b \citep{kamath2025gemma}, gemma3\_27b \citep{kamath2025gemma}, llama3.2\_3b \citep{grattafiori2024llama}, llama3.3\_70b \citep{grattafiori2024llama}, llama3\_70b \citep{grattafiori2024llama}, llama3\_8b \citep{grattafiori2024llama}. We select the following thinking models: gptoss\_20b \citep{agarwal2025gpt}, gptoss\_120b \citep{agarwal2025gpt}, qwen3\_1.7b \citep{yang2025qwen3}, qwen3\_4b \citep{yang2025qwen3}, qwen3\_8b \citep{yang2025qwen3} and qwen3\_32b \citep{yang2025qwen3}. We treat each round as an independent sample (more explanation in Appendix Section \ref{independent_sample}), but the LLMs have memory to keep track of previous stages in the same round, thus allowing the LLM to mimic the progression of the actual experiment. So, our total N is 1023 (3 rounds for each of the 341 ``digital twins''). We use the temperature of 1.2 to increase the diversity in model responses. All selected models were open-source and run locally using Ollama to preserve privacy and prevent data security issues.

% \subsection{Tables}
\subsection{Evaluation}
\vspace{-0.5em}
\subsubsection{Analysis for stage 1}
We use different metrics to compare the LLM responses with human response to quantify whether LLMs can emulate human belief responses given the human persona.  

\textbf{KL Divergence. }To evaluate whether LLMs can emulate human belief distributions, we use KL divergence \citep{Kullback_Leibler_1951} between the distribution of the initial Likert ratings selected by the LLMs with the distribution of the initial Likert ratings selected by humans. This is shown in equation \ref{KL_div}, where $P$ is the probability mass function.

\vspace{-1em}
\begin{equation}
\label{KL_div}
D_{\mathrm{KL}}\!\left(P_{\mathrm{actual}} \,\|\, P_{\mathrm{LLM}}\right)
=
\sum_{Likert_{initial}=0}^{4}
P_{\mathrm{actual}}(Likert_{initial})\,
\log\left(
\frac{P_{\mathrm{actual}}(Likert_{initial})}
{P_{\mathrm{LLM}}(Likert_{initial})}
\right)
\end{equation}
\vspace{-0.25em}

\textbf{Wasserstein Distance. } To evaluate whether LLMs can emulate human belief distributions, we calculate Wasserstein Distance \citep{ruschendorf1985wasserstein} between the distribution of the initial Likert ratings selected by the LLMs with the distribution of the initial Likert ratings selected by humans. Using the SciPy implementation, the equation is shown in \ref{wasserstein}. $P$ is the probability mass function, and F is the cumulative distribution function.

\vspace{-1em}
\begin{equation}
\label{wasserstein}
W_1\!\left(P_{\mathrm{actual}}, P_{\mathrm{LLM}}\right)
=
\sum_{Likert_{initial}=0}^{4}
\left|
F_{\mathrm{actual}}(Likert_{initial})
-
F_{\mathrm{LLM}}(Likert_{initial})
\right|
\end{equation}
\vspace{-0.25em}

\textbf{Mean and Standard Deviation.} We calculate the mean and standard deviation of the initial Likert rating of LLMs and humans. This is used to characterize the central tendency and dispersion of their belief distributions.

\textbf{Mann-Whitney U Test.} We also use a Mann-Whitney U Test to quantify if there is a difference between the LLM response distribution and the human response distribution. 

\textbf{Correlation Metrics.} We calculate the Spearman correlation, $\rho$\_{s}, between the LLM's initial Likert rating and the human's initial Likert rating. This provides a more one-to-one comparison of how well LLMs emulate individual human preferences. We use a t-test to evaluate the significance of the results.

\subsubsection{Analysis for stage 2}
\textbf{Comparing Change in Belief.} As defined in prior works \citep{proma2025exploring}, we term belief change or ``belief update'' as the difference in Likert rating between stage 2 and stage 1. We calculate this belief change for both LLMs and humans, and compare the means and standard deviations to understand how the distribution differs. We calculate Spearman correlation between the belief change in LLM and in humans to see how well LLMs emulate human belief changes. We used a t-test to evaluate the significance of the results. 

\textbf{Social Influence.} In stage 2 of the original experiment, the participants are shown the responses of a few others in the network (i.e., peers, $j$), and the authors measure how participants' beliefs change on seeing others' responses \citep{proma2025exploring, proma2025personalized}. Following prior studies, we calculate the difference between participants' initial belief and the average of those they see in stage 2, and then calculate the correlation between that and their belief change. For this paper, we term it as the ``social influence''. We calculate this metric for both the LLMs and the individuals, where $i$ can be either human participant or the LLM, and $N_j$ refers to the total number of peers.  

\vspace{-1em}
\begin{equation}
\label{eq:social_influence}
    \rho_{\text{social}} = SpearmanCorr\!\left(\frac{1}{N_j}\sum_{j=1}^{N_j} {Likert}_{j_{initial}} - {Likert}_{i_{initial}},\;\;
{Likert}_{i_{updated}} - {Likert}_{i_{initial}}\right)
\end{equation}
\vspace{-0.25em}

\subsubsection{Analysis for stage 3}
\textbf{Follow Signal \citep{proma2025exploring}.} In stage 3 of the original experiment, the participants select those they would like to ``follow'' (i.e., see more of in subsequent rounds). Adapting prior studies, we define follow signal, $Follow$, as the mean ratings of those the participant/LLM selected in stage 3. Assuming $j$ denotes each followed peer, and $N_{f_j}$ is the total number of followed peers,

\vspace{-1em}
\begin{equation}
\label{eq:F_i}
    Follow = \frac{1}{N_{f_j}}\sum_{j=1}^{N_{f_j}} Likert_{j_{initial}}
\end{equation}
\vspace{-0.25em}

We also calculate the Spearman correlation between LLM Follow Signal and Human Follow Signal. These metrics show how well LLMs emulate human follow preferences.

\textbf{Belief Network Distance.}  We adapt the Belief Network Distance from prior studies \citep{proma2025exploring}. Here, Belief Network Distance is the absolute difference between the initial Likert value and the Follow Signal.

\begin{equation}
\label{eq:B_followed}
     B_{i_{followed}} = |Follow_{i} - Likert_{i_{initial}}|
\end{equation}

We calculate the mean Belief Network Distance and also the Spearman correlation between the individual Belief Network Distance and the corresponding LLM Belief Network Distance. These metrics show how well LLMs emulate the overall distribution and the individual homophilic tendencies in stage 3. 
\section{Results}
\vspace{-0.5em}

\begin{figure}[t]
    \centering
    \begin{minipage}[c]{0.49\textwidth}
        \centering
        \includegraphics[width=\linewidth]{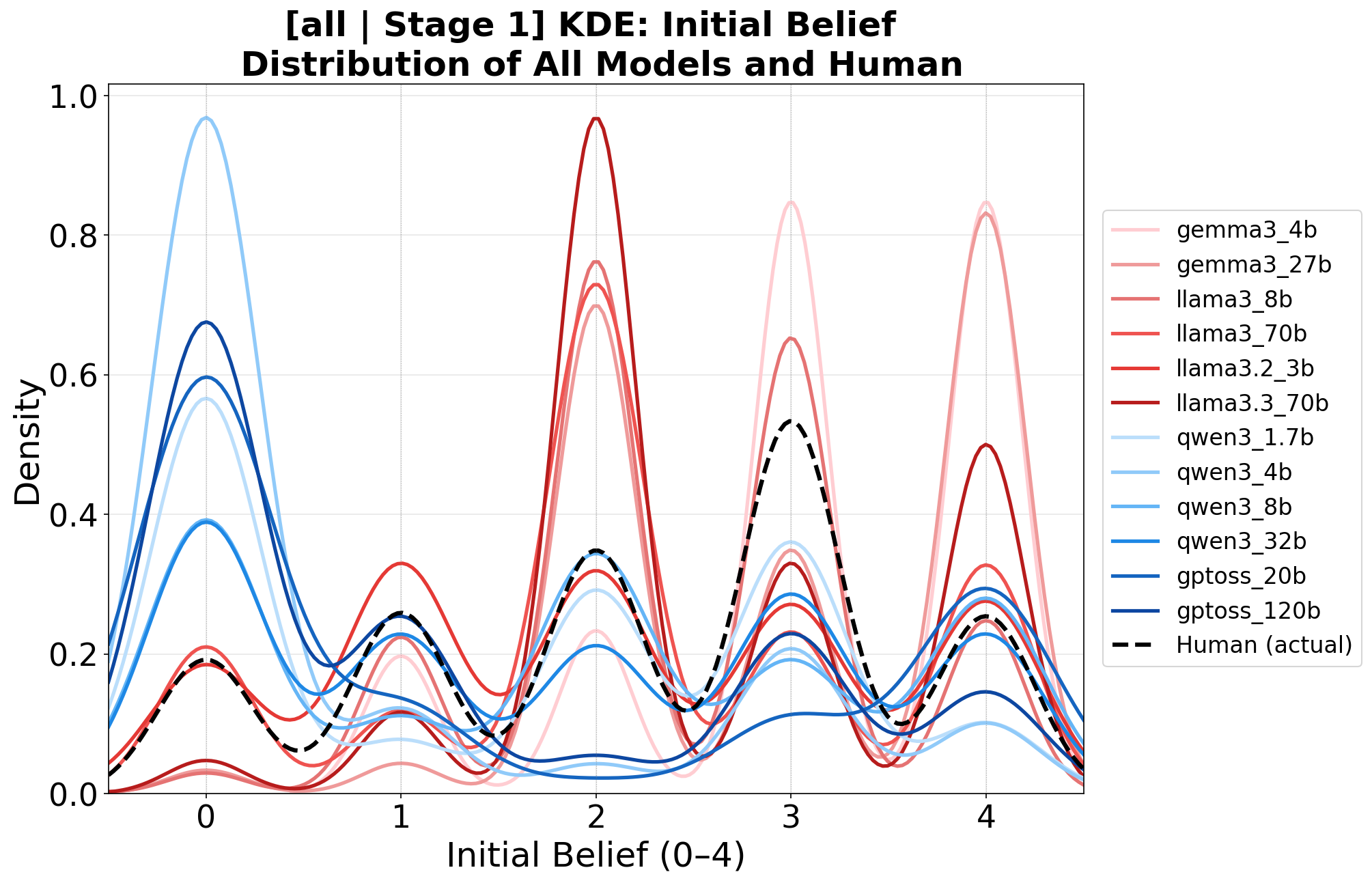}\par\vspace{0.5em}
        \includegraphics[width=\linewidth]{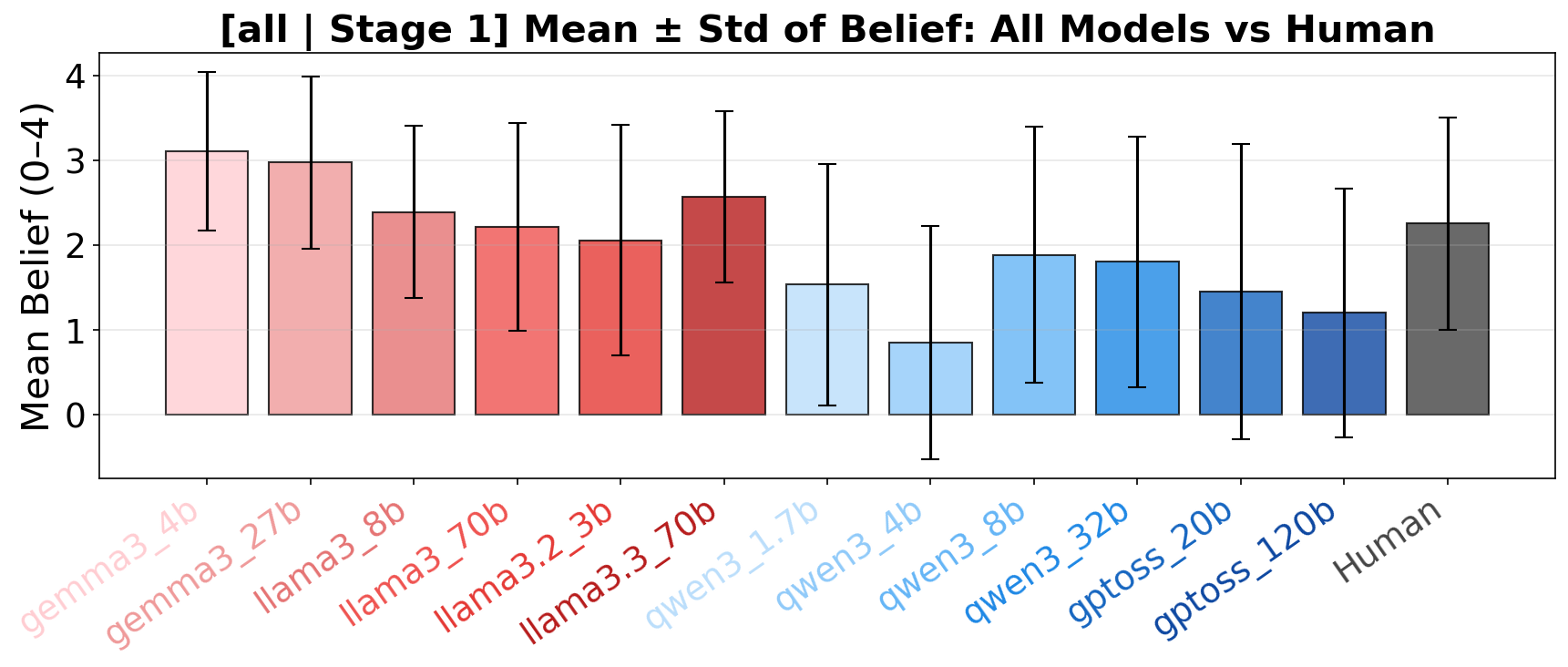}
    \end{minipage}
    \hfill
    \begin{minipage}[c]{0.48\textwidth}
        \centering
        \includegraphics[width=1.1\linewidth]{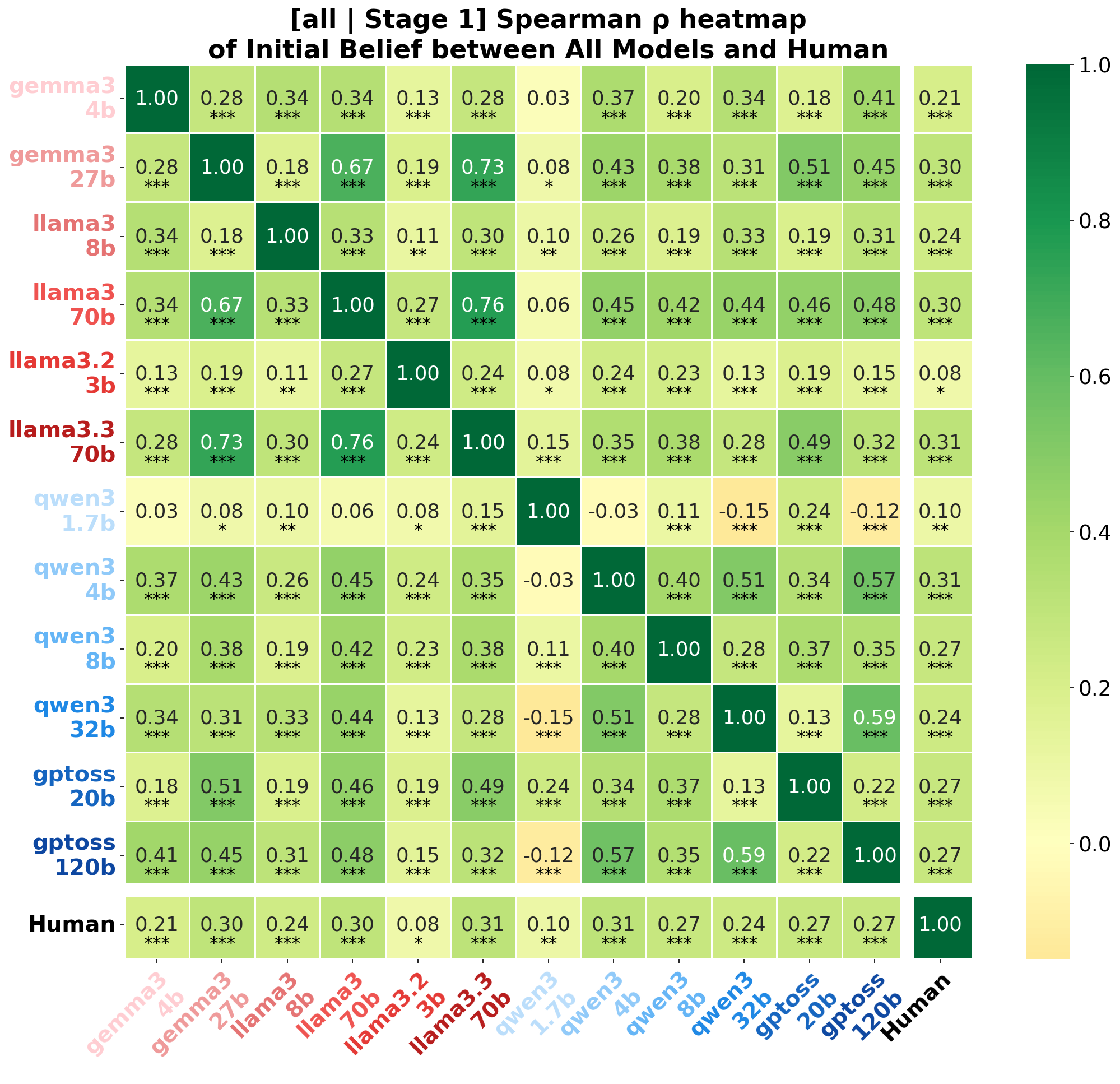}
    \end{minipage}
    \caption{The plots in the right show that non-thinking models tend to overestimate, and thinking models tend to underestimate initial Likert values. (a) Top-left: KDE of the initial Likert beliefs of various models and humans. (b) Bottom-left: Mean of the initial Likert Responses of various models and humans. Exact values are provided in Appendix, Table \ref{mean_std_s1}. (c) Right: Heatmap of Spearman correlation for initial Likert Responses. Shades of red represent the non-thinking models, and shades of blue represent the thinking models.}
    \label{fig:s1}
\end{figure}

\subsection{Finding 1: LLMs struggle to emulate initial human belief distributions}
\vspace{-0.5em}
In terms of overall distribution, KL divergence and Wasserstein Distance suggests that there is discrepancy between LLMs and human belief distributions (Table \ref{tab:s1}). As a baseline, we randomly split the human sample into two equal subgroups and calculated the KL divergence and Wasserstein Distance, with values of 0.006 and 0.064 respectively. Compared to this baseline, KL divergence and Wasserstein Distance values for LLM distributions are much larger (Table \ref{tab:s1}). Furthermore, Mann-Whitney U test shows that the LLM distributions and human distributions are significantly different 10 out of 12 times. 

The response patterns also vary across different models. The KDE plot shows that most non-thinking models peak at 2, while all thinking models peak at 0 (Figure \ref{fig:s1}a). This suggests that thinking models tend to disagree with the given statement more. Similarly, considering the mean belief of LLMs compared to humans, non-thinking models overestimate and thinking models underestimate the average mean belief (Figure \ref{fig:s1}b, Table \ref{mean_std_s1}). 

We use Spearman correlation between initial Likert values provided by LLMs and humans for a one-on-one comparison between LLMs and humans. There is moderate positive correlation, with the highest correlation between an LLM (llama3.3\_70b) and human being 0.31 (Figure \ref{fig:s1}c). Spearman correlation across models varies. Therefore, it is evident that there are certain limitations in LLM emulation of humans' initial beliefs. 

\begin{table}[t]
\centering
\small{
\begin{tabular}{lllll}
\hline
Model Type   & Model         & KL Divergence & Wasserstein Dist & \shortstack{Mann-Whitney\\ U p-value} \\
\hline
Non-Thinking & gemma3\_4b    & 2.5683        & 0.8520           & 0.0000***             \\
Non-Thinking & gemma3\_27b   & 0.5121        & 0.7200           & 0.0000***             \\
Non-Thinking & llama3\_8b    & 0.2030        & 0.3102           & 0.1219                \\
Non-Thinking & llama3\_70b   & 0.2133        & 0.2830           & 0.1669                \\
Non-Thinking & llama3.2\_3b  & 0.0850        & 0.3483           & 0.0000***             \\
Non-Thinking & llama3.3\_70b & 0.3402        & 0.4653           & 0.0000***             \\
\hline
& Average       &  0.6537             & 0.4965           &                 \\
\hline
Thinking     & qwen3\_1.7b   & 0.2595        & 0.7180           & 0.0000***             \\
Thinking     & qwen3\_4b     & 0.7593        & 1.3955           & 0.0000***             \\
Thinking     & qwen3\_8b     & 0.2021        & 0.4723           & 0.0000***             \\
Thinking     & qwen3\_32b    & 0.1054        & 0.4713           & 0.0000***             \\
Thinking     & gptoss\_20b   & 0.8100        & 0.9929           & 0.0000***             \\
Thinking     & gptoss\_120b  & 0.4846        & 1.0566           & 0.0000***  \\
\hline
     & Average       &   0.4368          & 0.8511           &  
\\ \hline
\end{tabular}
}
\caption{KL Divergence, Wasserstein Distance, and Mann-Whitney U p-values to compare the LLM belief distribution with the human belief distribution for stage 1. Significance: * = p \textless 0.05, ** = p \textless 0.01, *** = p \textless 0.001}
\label{tab:s1}
\end{table}

\subsection{Finding 2: LLMs change their responses to be closer to their  ``social influence''}
\begin{figure}[t]
    \centering
    \includegraphics[width=0.8\linewidth]{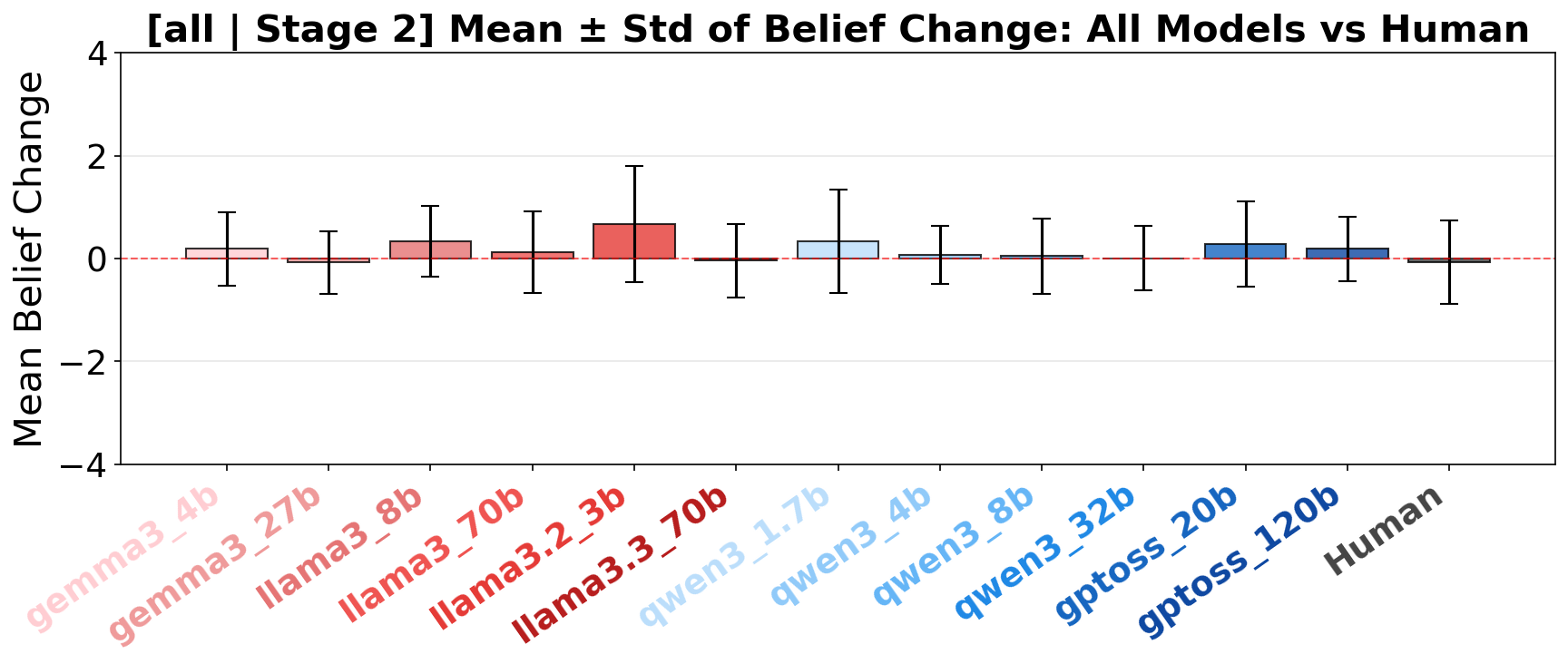}
    \par\vspace{0.5em}
    \includegraphics[width=0.9\linewidth]{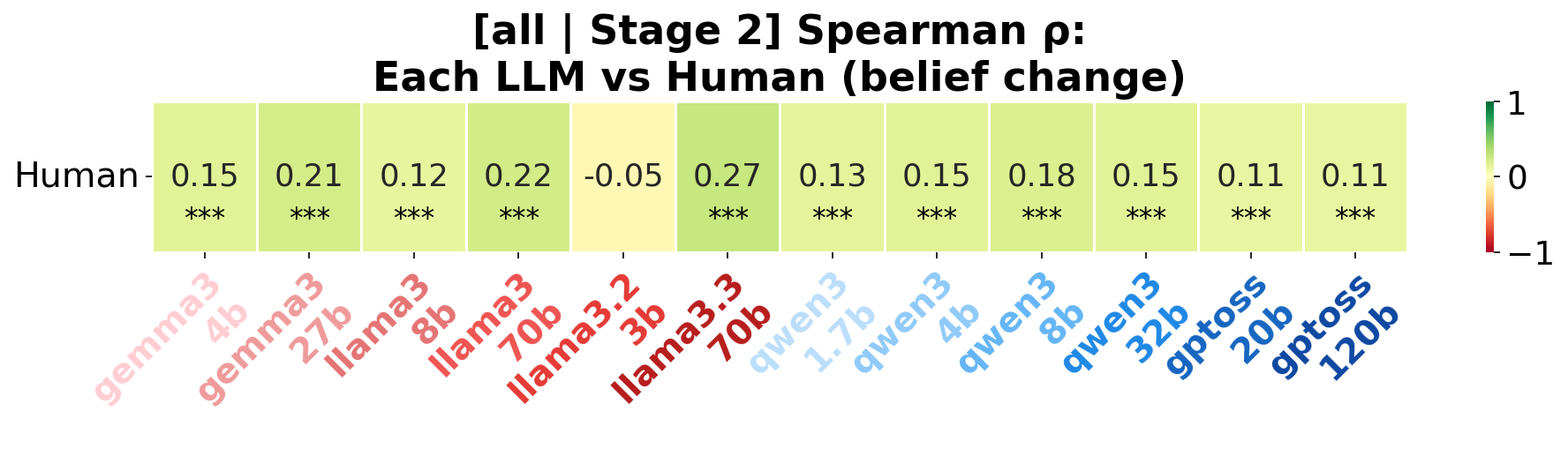}
    \caption{Stage 2 belief change analysis. (a) Top: Mean and standard deviation of belief change across all models and human participants. (b) Bottom: Spearman correlation between each LLM's belief change and human belief change.}
    \label{fig:s2}
\end{figure}

\vspace{-0.5em}
Comparing mean changes in beliefs for humans and LLMs, humans tend to change their beliefs less compared to LLMs (Figure \ref{fig:s2}a), i.e., humans are more rigid in their beliefs. There is weak Spearman correlation between how LLMs change their beliefs and how individuals change their beliefs (Figure \ref{fig:s2}b). 

%Note that there is no difference between a negative belief change and a positive belief change in this case; only the extent of the change matters. 

Following prior studies \citep{proma2025exploring, proma2025personalized}, we calculate the difference between participants' initial belief and the average of those they see in stage 2, and then calculate the correlation between that and their belief change. The ``social influence'' (\ref{eq:social_influence}) is overall higher for LLMs compared to humans (11/12 times) (Table \ref{tab:social_influence}), with significant Fisher r-to-z p values 8 out of those 11 times. This suggests that LLMs are more conforming than humans. 

\begin{table}[t]
\small{
\centering 
\begin{tabular}{lllll}
\hline
\shortstack{Model\\Type} & Model & \shortstack{Actual Social\\Influence (Spearman)} & \shortstack{LLM Social\\Influence (Spearman)} & \shortstack{Fisher\\p-value} \\
\hline
Non-Thinking & gemma3\_4b    & 0.3215*** & 0.4948*** & 0.0000*** \\
Non-Thinking & gemma3\_27b   & 0.3215*** & 0.3447*** & 0.5615    \\
Non-Thinking & llama3\_8b    & 0.3202*** & 0.5049*** & 0.0000*** \\
Non-Thinking & llama3\_70b   & 0.3215*** & 0.4899*** & 0.0000*** \\
Non-Thinking & llama3.2\_3b  & 0.3438*** & 0.5511*** & 0.0000*** \\
Non-Thinking & llama3.3\_70b & 0.3215*** & 0.4379*** & 0.0024**  \\
\hline
 & Average       & 0.3250    & 0.4706    & 0.0940    \\
\hline
Thinking     & qwen3\_1.7b   & 0.3219*** & 0.5017*** & 0.0000*** \\
Thinking     & qwen3\_4b     & 0.3265*** & -0.0221   & 0.0000*** \\
Thinking     & qwen3\_8b     & 0.3215*** & 0.3770*** & 0.1605    \\
Thinking     & qwen3\_32b    & 0.3215*** & 0.4471*** & 0.0010**  \\
Thinking     & gptoss\_20b   & 0.3221*** & 0.5284*** & 0.0000*** \\
Thinking     & gptoss\_120b  & 0.3235*** & 0.3833*** & 0.1289    \\
\hline
    & Average       & 0.3228    & 0.3692    & 0.0484   
\\
\hline
\end{tabular}
}
\caption{``Social Influence'' (Equation \ref{eq:social_influence}) in stage 2, which represents a Spearman correlation metric measuring how others' opinions influence response change. Actual Social Influence (Spearman) shows that metric for participants, and LLM Social Influence (Spearman) shows that for LLMs. If an LLM fails to generate a certain instance, we drop that instance from the human data as well during comparison with that specific LLM, resulting in slight variations in actual social influence. Fisher p-value shows if the LLM correlation is significantly different from the human correlations. Significance: * = p \textless 0.05, ** = p \textless 0.01, *** = p \textless 0.001}
\label{tab:social_influence}
\end{table}

\subsection{Finding 3: LLMs take a nuanced approach to emulating human selections and homophilic behavior}
\begin{figure}[t]
    \centering
    \begin{minipage}[c]{0.49\textwidth}
        \centering
        \includegraphics[width=\linewidth]{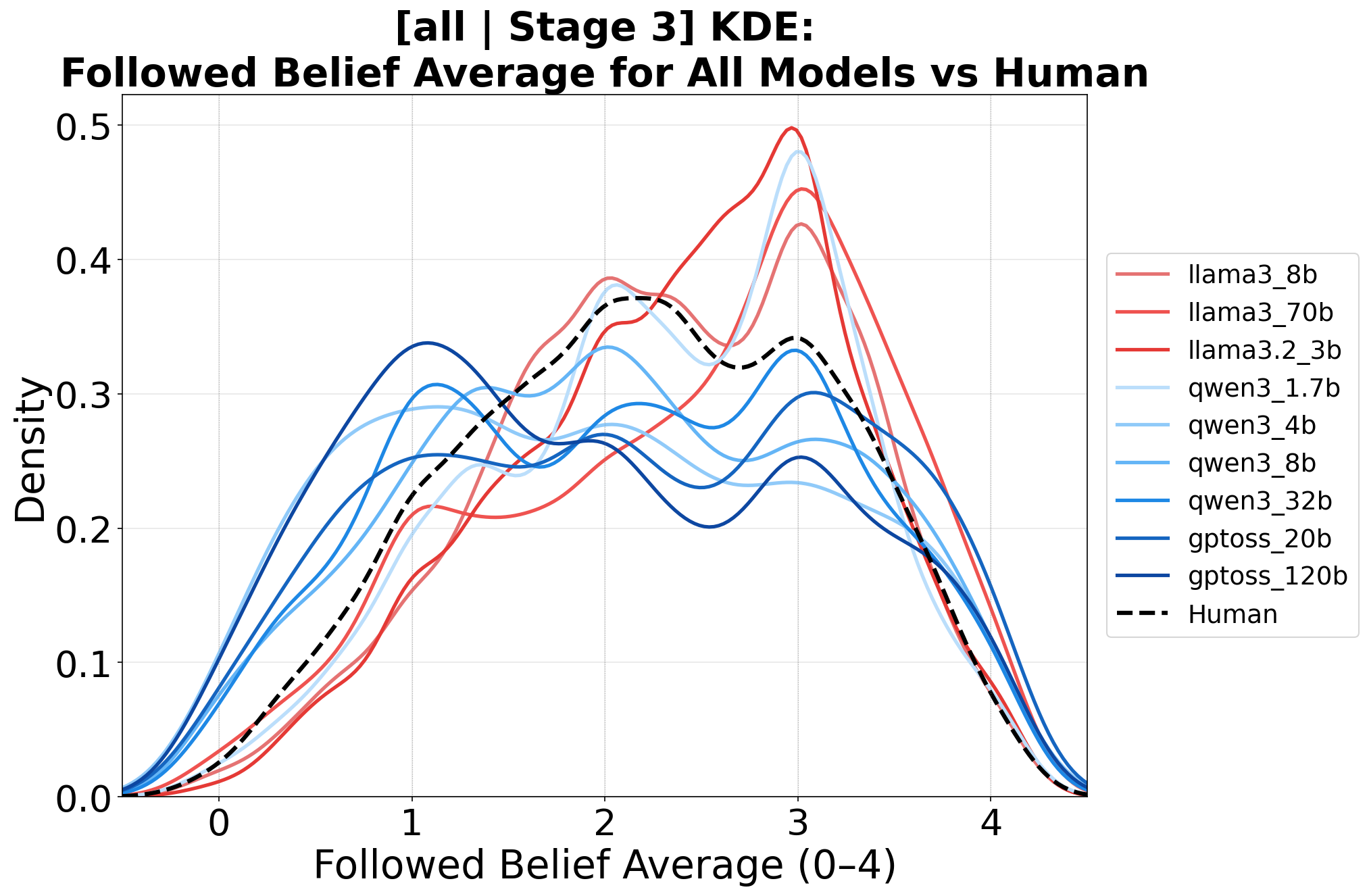} \par\vspace{0.5em}
        \includegraphics[width=\linewidth]{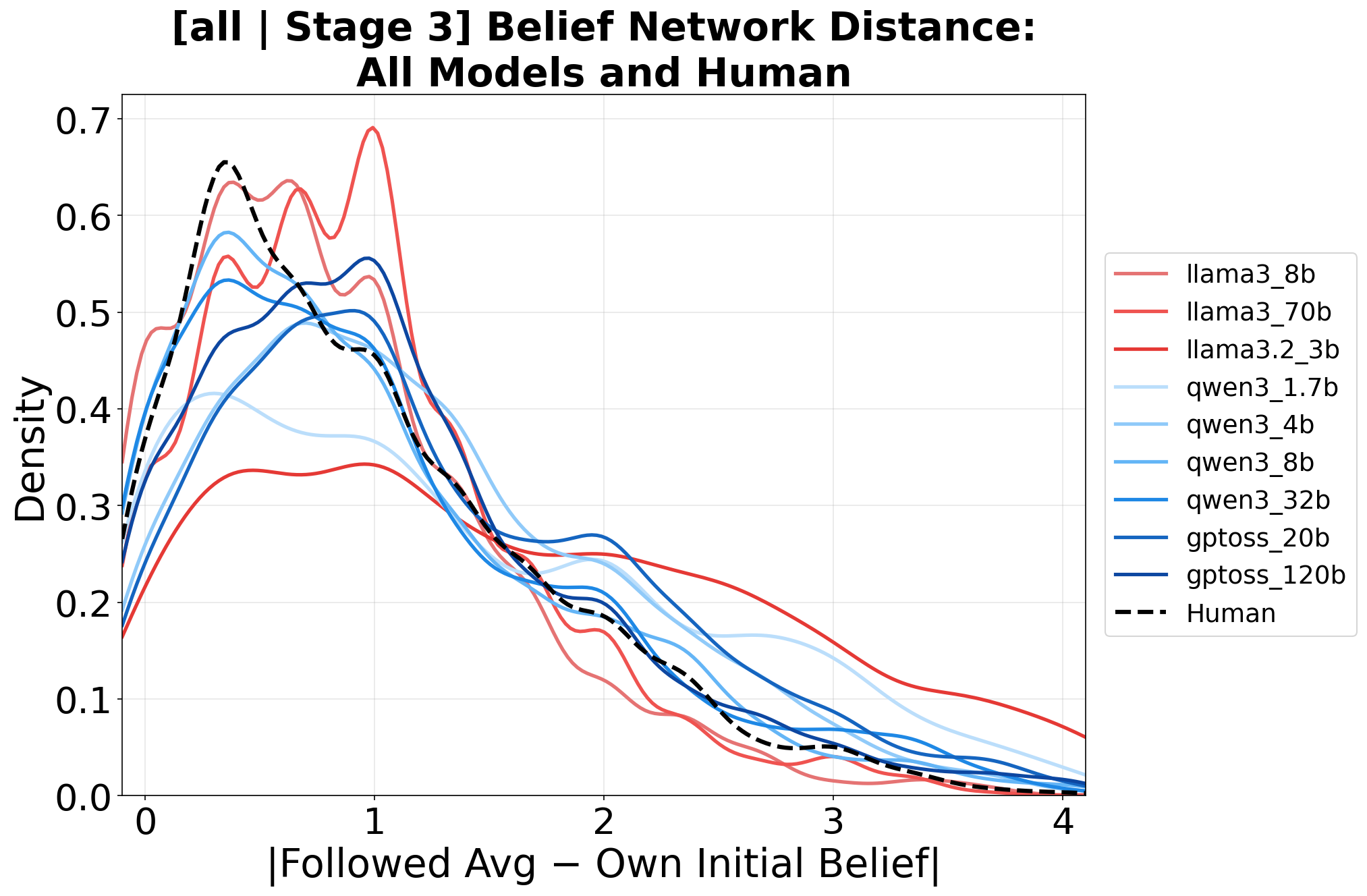}
    \end{minipage}
    \hfill
    \begin{minipage}[c]{0.48\textwidth}
        \centering
        \includegraphics[width=1.1\linewidth]{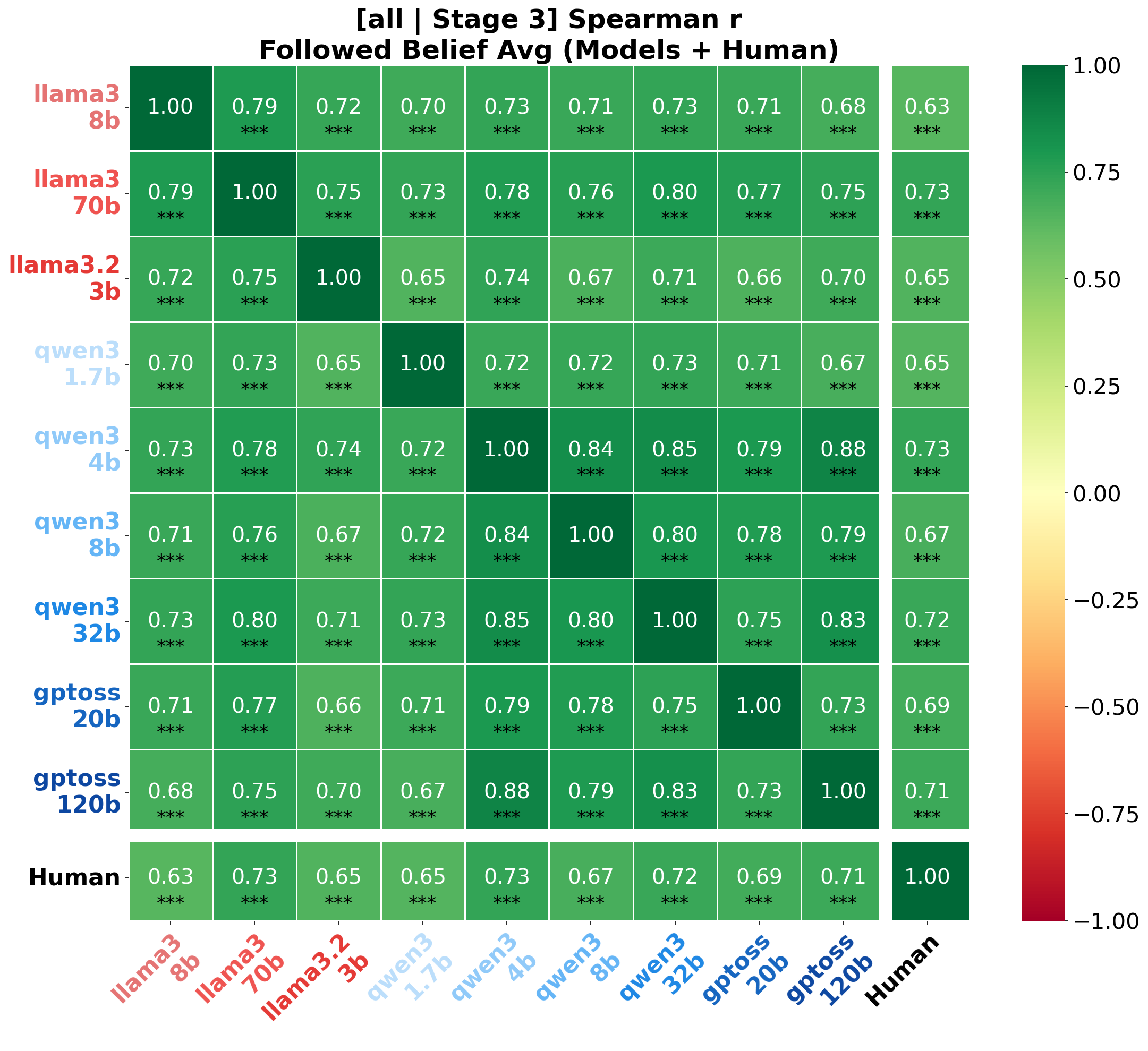}
    \end{minipage}
    \caption{(a) Top-left: KDE distribution of Follow Signal (b) Bottom-left: KDE distribution of Belief Network Distance (Equation \ref{eq:B_followed}) for all models and human participants in Stage 3. (c) Right: Spearman Correlation between models and humans for Follow signal. Please note that the JSON format required for this stage emulation was not supported by some of the non-thinking models (gemma3\_27b, gemma3\_4b and llama3.3\_70b), and hence they have been dropped for stage 3 analyses.}
    \label{fig:s3}
\end{figure}

\vspace{-0.5em}
We calculate the Follow Signal (Equation \ref{eq:F_i}) for both LLMs and humans, and plot the KDE distributions in Figure \ref{fig:s3}a. The KDE suggests that there are similarities between the distributions of how opinions are selected by humans and LLMs. This is further confirmed by the high correlation between the Follow signal for both, as shown in Figure \ref{fig:s3}c, suggesting LLMs are quite good at simulating what humans might select in their networks. Cross-modal analysis of Spearman correlation suggests that LLMs are highly correlated with each other as well (Figure \ref{fig:s3}c). 

Next, we use the Belief Network Distance to measure homophily tendency, since a key aspect of homophily is that it is relative to one's own preferences. The Belief Network Distance measures how close the responses selected are to the humans/LLMs' initial responses (Figure \ref{fig:s3}b, Table \ref{tab:BeliefNetwork}). Comparing the mean LLM Belief Network Distance and human Belief Network Distance, LLM means are consistently higher, suggesting that they tend to select peers who are slightly more distant from their own beliefs than humans do. Based on the KDE plot and Mann-Whitney U tests, the distributions of Belief Network Distance differ between LLMs and humans. This suggests that although LLMs can emulate human choices, they cannot emulate the homophilic nature of humans, i.e., their selections are not based as strongly as humans on their prior beliefs/actions. 

\begin{table}[t]
\small{
\begin{tabular}{lllll}
\hline 
Model Type &
  Model &
  \begin{tabular}[c]{@{}l@{}}Mean LLM Belief \\ Network Distance\end{tabular} &
  \begin{tabular}[c]{@{}l@{}}Mean Human Belief \\ Network Distance\end{tabular} &
  \begin{tabular}[c]{@{}l@{}}MWU p (Human \\ Belief  Network \\ Distance)\end{tabular} \\
\hline 
Non-Thinking & gemma3\_4b    & N/A                 & 0.9530 (sd: 0.7705) & N/A       \\
Non-Thinking & gemma3\_27b   & N/A                 & 0.9530 (sd: 0.7705) & N/A       \\
Non-Thinking & llama3\_8b    & 0.8554 (sd: 0.7130) & 0.9530 (sd: 0.7705) & 0.0140*   \\
Non-Thinking & llama3\_70b   & 0.9491 (sd: 0.6851) & 0.9530 (sd: 0.7705) & 0.1726    \\
Non-Thinking & llama3.2\_3b  & 1.6004 (sd: 1.1442) & 0.9530 (sd: 0.7705) & 0.0000*** \\
Non-Thinking & llama3.3\_70b & N/A                 & 0.9530 (sd: 0.7705) & N/A       \\ \hline 
Non-Thinking & Average       & 1.1350              & 0.9530              &           \\ \hline 
Thinking     & qwen3\_1.7b   & 1.3048 (sd: 1.0408) & 0.9530 (sd: 0.7705) & 0.0000*** \\
Thinking     & qwen3\_4b     & 1.1960 (sd: 0.8634) & 0.9530 (sd: 0.7705) & 0.0000*** \\
Thinking     & qwen3\_8b     & 1.0033 (sd: 0.8501) & 0.9530 (sd: 0.7705) & 0.4360    \\
Thinking     & qwen3\_32b    & 1.0375 (sd: 0.8750) & 0.9530 (sd: 0.7705) & 0.1628    \\
Thinking     & gptoss\_20b   & 1.2412 (sd: 0.8911) & 0.9530 (sd: 0.7705) & 0.0000*** \\
Thinking     & gptoss\_120b  & 1.0674 (sd: 0.8245) & 0.9530 (sd: 0.7705) & 0.0014**  \\ \hline 
Thinking     & Average       & 1.1417              & 0.9530              &   
\\ \hline 
\end{tabular}
}
\caption{Statistics comparing responses of different LLM models to the actual human responses for stage 3. Belief Network Distance signifies how far participants' / LLMs' initial beliefs are to who they select for their networks. Please note that the JSON format required for this stage emulation was not supported by gemma3\_27b, gemma3\_4b and llama3.3\_70b (hence N/A for those models). Significance: * = p \textless 0.05, ** = p \textless 0.01, *** = p \textless 0.001}
\label{tab:BeliefNetwork}
\end{table}

% \begin{figure}[t]
%     \centering
%     \includegraphics[width=0.8\linewidth]{s3_gap_kde.png}
%     \caption{KDE distribution of Belief Network Distance (Equation \ref{eq:B_followed}) for all models and human participants in Stage 3. Most LLM distributions closely overlap with the human distribution, suggesting that LLMs approximate human homophilic tendencies in network selection, though several models exhibit slightly heavier right tails indicating occasional selection of more distant opinions.}
%     \label{fig:BeliefNetwork}
% \end{figure}

\section{Discussion, limitations, and future work}
\textbf{Implications.} Our findings also contribute to the broader research question of understanding the capabilities of LLMs in simulations. We show that LLM behavior diverges from humans when it comes to capturing belief distributions, belief changes and network restructuring. Several factors may explain the differences between thinking and non-thinking models in initial belief distributions. For example, 2/3 of the statements are false, which thinking models may perform better at identifying due to their reasoning capabilities or knowledge cutoff dates (see Appendix, Table \ref{tab:model_cutoffs}). When shown the opinions of others, LLMs are more easily influenced compared to humans in changing their responses, consistent with existing literature suggesting LLMs tend to be more agreeable and sycophant with human opinions, often even going against ground truth \citep{sharma2023towards, clark2025epistemic, wang2026truth, cheng2025social}. While LLMs emulate human ``follow'' preferences, they show higher Belief Network Distance on average. This suggests that they struggle with capturing the extent of homophily in human networks. As there is a push toward using LLM-based agents to simulate large-scale social systems \citep{Park_OBrien_Cai_Morris_Liang_Bernstein_2023}, our findings highlight potential pitfalls and can inform future research on where LLMs can and cannot be used as human proxies.  

% \textcolor{red}{In terms of belief distributions, we hypothesize that since 2/3 of the statements in the experiment are false, thinking models may be better at identifying that, while non-thinking models initially tend to be more agreeable.} 

%Prior works on vaccine simulation show that LLMs can simulate aspects of human behavior, with certain models doing better than others, but there are still many real-world alignment issues \citep{hou2025can}, and that is what we see in our work as well. 
% TO DO: DISCUSSION on why if s1 is so bad, how are the rest of it doing better? What might it mean for LLMs for such sequential tasks? 

%TO DO, benchmark against some other papers doing simulation. what were the metrics used, and what did they consider to be "good"

\textbf{Our results with respect to LLM scaling laws.} As we evaluate models of various parameters, we note that models with larger parameters do not necessarily always perform better, which challenge the traditional scaling law expectations that increasing model scale would improve alignment to human benchmarks. One explanation could be that the nature of belief rigidity is inherently different from tasks LLMs are trained for like reasoning, knowledge retrieval, and instruction following, where the scaling laws are more applicable. Emulating belief dynamics requires reproducing various human tendencies and psychological processes (such as motivated reasoning \citep{pate2026replicating}), which LLMs might struggle with. 

\textbf{Design choices and future work.} In our experiments, the LLMs are provided with a short-term memory where they remember their actions for the previous stages of the same round. However, they do not have memory across rounds, and each round is treated as an independent sample. This design choice ensured that the information provided was within the context window of the models. Moreover, we include only demographic information and the Big-5 personality traits as the persona of the model. While prior works show that elaborate LLMs personas tend to be more accurate in simulations \citep{park2024generative}, it may not always be possible to have access to that level of information. Our work therefore still highlights valuable findings related to LLM simulations, specifically that the ``digital twin'' method may not work with limited persona information. Additionally, we used only open source models for data protection and privacy reasons. Future work can, therefore, include adding memory across rounds, and also experimenting with more elaborate personas, such as including information related to social media habits, news habits, and lived experiences. Additionally, researchers should also test closed-source models, which might have stronger performance as emulators. 

Our findings underscore a fundamental asymmetry, which is that LLMs process social context differently from humans. This gap shows up in various ways during simulations, especially when simulating belief dynamics. Human belief dynamics are complex processes shaped through identity, motivation, and affect, and further work is needed to better align LLMs to humans if they are to be used for simulations.

\section*{Ethics statement}
To preserve the privacy of the human data used, we used only open-sourced models that can be locally hosted in the university server. This ensures that there is no breach of data privacy issues. 

LLMs were used as writing assistant and for grammar checks, which is in line with COLM's policies. All content was checked and verified by the authors.
% Authors can add an optional ethics statement to the paper. 
% For papers that touch on ethical issues, this section will be evaluated as part of the review process. The ethics statement should come at the end of the paper. It does not count toward the page limit, but should not be more than 1 page. 

\section*{Reproducibility statement}
For the analyses conducted, we reached out to the authors of the original paper \citep{proma2025personalized}, who then provided us access to the subset of the data used. Other researchers, if interested in this work or the dataset, are encouraged to reach out to the original authors \citep{proma2025personalized}. In terms of model simulation, we provide the exact models and the parameters used so that other researchers can reproduce the work. All models are open-sourced.

%\section*{LLM use disclosure}

\bibliography{ref}
\bibliographystyle{colm2026_conference}

\appendix

\section{Prompts used for simulation}
\label{prompts}
The prompts used for simulation are provided below for each stage. For the analysis, \texttt{memory\_summary} is empty and is passed as an empty string.

\subsubsection*{Stage 1}

\begin{lstlisting}
You are the person agent_id={agent_id}. You have the following persona={persona}.
Rate the following statement on a Likert scale from 0 to 4 based on how much you believe the statement to be true,
where 0 = strongly disagree and 4 = strongly agree.

Statement: "{statement}"

Format your response EXACTLY as follows:
Rating: <number>
Reason: <a short paragraph>
\end{lstlisting}

\subsubsection*{Stage 2}

\begin{lstlisting}
You are the person agent_id={self_profile.get("agent_id")}. You have the following persona: {self_profile.get("persona")}.
- Memory: {self_profile.get("memory_summary")}

The statement to evaluate:
"{statement}"

Your current belief about the statement is {prior} because: {self_profile.get("initial_rationale")}.
Now you observe the following neighbor ratings and rationales: {obs_str}

Task:
Considering your own persona and your observed neighbor opinions, rate how much you believe the given statement to be true on a Likert scale from 0 to 4, where:
0 = strongly disagree; 4 = strongly agree

Provide a short rationale (1--2 sentences) explaining your update.

Format your response EXACTLY as:
Rating: <number>
Reason: <a short paragraph>
\end{lstlisting}

Here, \texttt{\{obs\_str\}} is a list of the Likert responses and reasons provided by others for the same statement.

\subsubsection*{Stage 3}

\begin{lstlisting}
You are a person with the following profile.
- id: {self_profile.get("agent_id")}
- name: {self_profile.get("name")}
- persona: {self_profile.get("persona")}
- memory: {self_profile.get("memory_summary")}

The statement to evaluate: "{statement}"

Here is what you did in the previous stages:
{self_profile.get("stage1_summary", "")}
{self_profile.get("stage2_summary", "")}

You must choose exactly {k} candidates from the list below.
Candidates: {candidates_block}

Respond ONLY in strict JSON:
{
  "follow_ids": ["<id1>", "<id2>", ... exactly {k} ids ...],
  "reason": "<short explanation>"
}
\end{lstlisting}

Here, \texttt{\{candidates\_block\}} is a list of the Likert responses and reasons provided by potential candidates from whom the LLM must make its selections.

\section{Treating each round as an independent sample}
\label{independent_sample}
While each of the participants completed three rounds sequentially in the actual experiment, in our simulation, each round is treated as an independent sample. This design choice ensures that the prompts are within the context window of all the different LLM models with different parameter sizes tested in the simulation. Adding every single prior behavior would exponentially increase the prompt size as the experiment progresses, and smaller models may not be able to process it. Secondly, it prevents cascading of errors by the LLMs. In scenarios where the LLM does poorly at emulating in one of the rounds, our design ensures that the error does not impact future rounds.  

\section{Additional Results}

\subsection{Mean and Standard deviation of initial Likert beliefs of various models compared to humans}
The means and standard deviations for initial Likert belief ratings are provided at Table \ref{mean_std_s1}. 

\begin{table}[h]
\begin{tabular}{llllll}
Model Type   & Model         & LLM Mean & LLM Std & Actual Mean & Actual Std \\
\hline
Non-Thinking & gemma3\_4b    & 3.1037   & 0.9319  & 2.2518      & 1.2490     \\
Non-Thinking & gemma3\_27b   & 2.9718   & 1.0121  & 2.2518      & 1.2490     \\
Non-Thinking & llama3\_8b    & 2.3887   & 1.0145  & 2.2518      & 1.2490     \\
Non-Thinking & llama3\_70b   & 2.2145   & 1.2247  & 2.2518      & 1.2490     \\
Non-Thinking & llama3.2\_3b  & 2.0578   & 1.3569  & 2.3470      & 1.2159     \\
Non-Thinking & llama3.3\_70b & 2.5700   & 1.0092  & 2.2518      & 1.2490     \\
\hline
Non-Thinking & Average       & 2.5511   & 1.0915  & 2.2676      & 1.2435     \\
\hline
Thinking     & qwen3\_1.7b   & 1.5337   & 1.4203  & 2.2518      & 1.2490     \\
Thinking     & qwen3\_4b     & 0.8550   & 1.3736  & 2.2505      & 1.2494     \\
Thinking     & qwen3\_8b     & 1.8842   & 1.5078  & 2.2518      & 1.2490     \\
Thinking     & qwen3\_32b    & 1.8006   & 1.4805  & 2.2518      & 1.2490     \\
Thinking     & gptoss\_20b   & 1.4531   & 1.7375  & 2.2543      & 1.2482     \\
Thinking     & gptoss\_120b  & 1.2002   & 1.4621  & 2.2568      & 1.2465     \\
\hline
Thinking     & Average       & 1.4545   & 1.4970  & 2.2528      & 1.2485 
\\ \hline
\end{tabular}
\caption{Mean and standard deviation for each of the models' and the humans' initial Likert ratings. Note that in case an LLM fails to generate a certain instance, we drop that instance from the human data as well during comparing with that specific LLM. That is why there is slight variation in the mean and standard deviation columns for humans.}
\label{mean_std_s1}
\end{table}

\subsection{There is weak correlation between how different models change belief in stage 2}
\begin{figure}[t]
    \centering
    \includegraphics[width=0.9\linewidth]{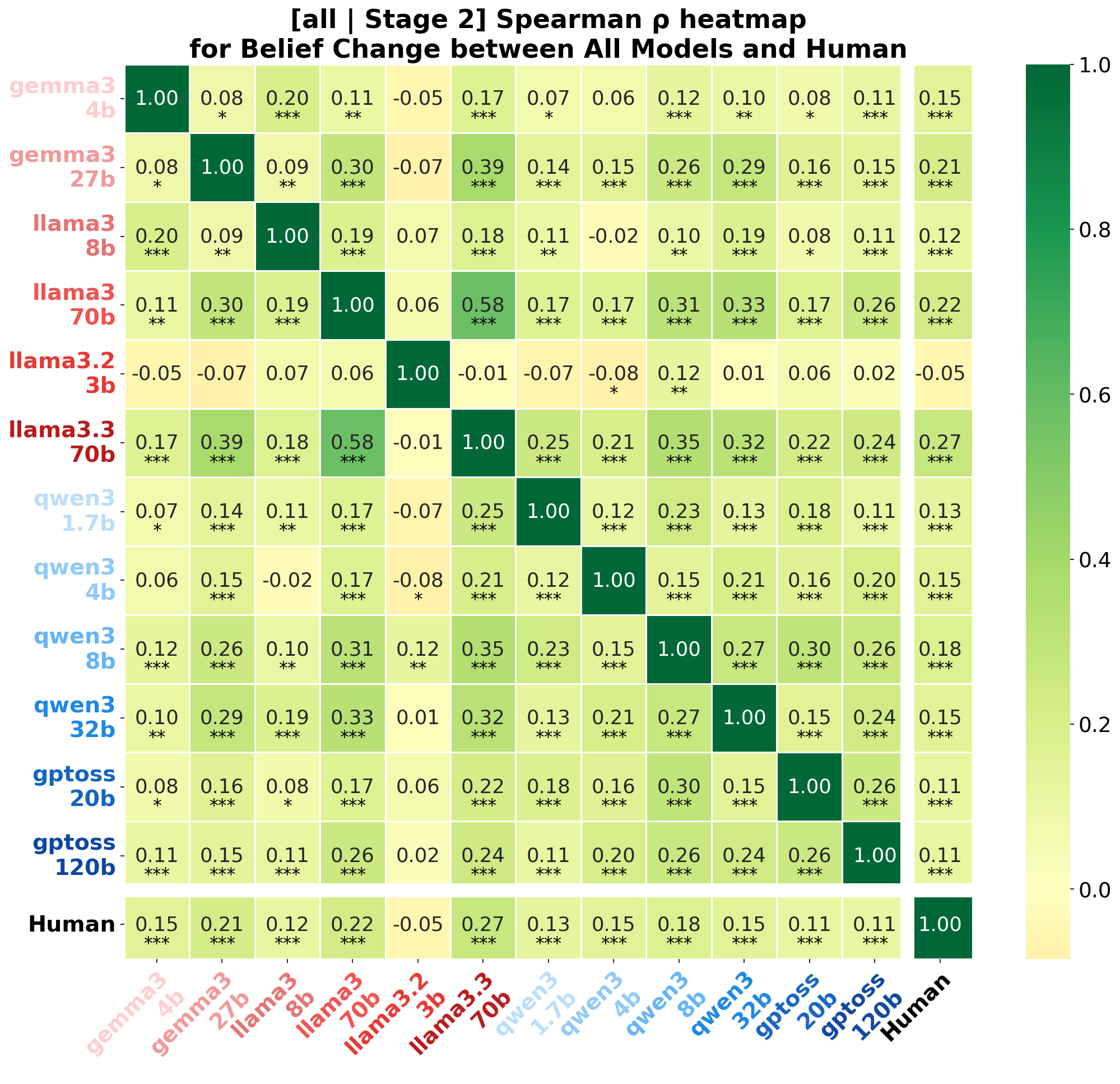}
    \caption{Spearman correlation heatmap showing belief change across various models and human.}
    \label{fig:cross_modal_belief_s2}
\end{figure}

Across models, there is no clear pattern on how various models change their beliefs in stage 2. This is shown in Figure \ref{fig:cross_modal_belief_s2}.

% \section{Additional Results}
% To analyze whether LLM simulation behavior is dependent on the topics being discussed, we analyze LLM simulation for each of the topics individually (Tables \ref{tab:s1_topicwise}, \ref{tab:s2_topicwise}, and \ref{tab:s3_topicwise}). Our findings suggest that models do not possess a generalized, topic-agnostic framework for social simulation, and that their behavior in simulations is contextual. 

\section{Knowledge Cutoff Dates for each Model }

\begin{table}[h]
\centering
\renewcommand{\arraystretch}{1.2}
\begin{tabular}{lll}
\hline
\textbf{Model} & \textbf{Training Cutoff Date} & \textbf{Source}\\ \hline
gemma3\_4b & March 2024 & https://gradientflow.com/gemma-3-what-you-need-to-know/ \\ \hline
gemma3\_27b & March 2024 & https://gradientflow.com/gemma-3-what-you-need-to-know/\\ \hline
llama3\_8b & March 2023 & https://huggingface.co/meta-llama/Meta-Llama-3-8B \\ \hline
llama3\_70b & December 2023 & https://huggingface.co/meta-llama/Meta-Llama-3-8B\\ \hline
llama3.2\_3b & December 2023 & https://huggingface.co/meta-llama/Llama-3.2-3B\\ \hline
llama3.3\_70b & December 2023 & https://huggingface.co/meta-llama/Llama-3.3-70B-Instruct\\ \hline
qwen3\_1.7b & Not Disclosed &\\ \hline
qwen3\_4b & Not Disclosed &\\ \hline
qwen3\_8b & Not Disclosed &\\ \hline
qwen3\_32b & Not Disclosed &\\ \hline
gptoss\_20b & June 2024 & https://openai.com/index/gpt-oss-model-card/\\ \hline
gptoss\_120b & June 2024 & https://openai.com/index/gpt-oss-model-card/\\ \hline
\end{tabular}
\caption{Training Cutoff Dates for Evaluated LLMs}
\label{tab:model_cutoffs}
\end{table}

% \bibliography{reflinks}
% \bibliographystyle{colm2026_conference}

\end{document}